\documentclass[twoside,twocolumn,9pt]{article}
\usepackage{extsizes}
\usepackage{amsmath,amssymb}
\usepackage[super,sort&compress,comma]{natbib} 
\usepackage[version=3]{mhchem}
\usepackage[left=1.5cm, right=1.5cm, top=1.785cm, bottom=2.0cm]{geometry}
\usepackage{balance}
\usepackage{bm}
\usepackage{times,mathptmx}
\usepackage{sectsty}
\usepackage{graphicx} 
\usepackage{lastpage}
\usepackage[format=plain,justification=justified,singlelinecheck=false,font={stretch=1.125,small,sf},labelfont=bf,labelsep=space]{caption}
\usepackage{float}
\usepackage{fancyhdr}
\usepackage{fnpos}
\usepackage[english]{babel}
\addto{\captionsenglish}{%
  
}
\usepackage{array}
\usepackage{color}

\usepackage{droidsans}
\usepackage{charter}
\usepackage[T1]{fontenc}
\usepackage[usenames,dvipsnames]{xcolor}
\usepackage{setspace}
\usepackage[compact]{titlesec}
\usepackage{hyperref}

\usepackage{epstopdf}

\definecolor{cream}{RGB}{222,217,201}

\begin{document}

\pagestyle{fancy}
\thispagestyle{plain}
\fancypagestyle{plain}{
\fancyhead[C]{\includegraphics[width=18.5cm]{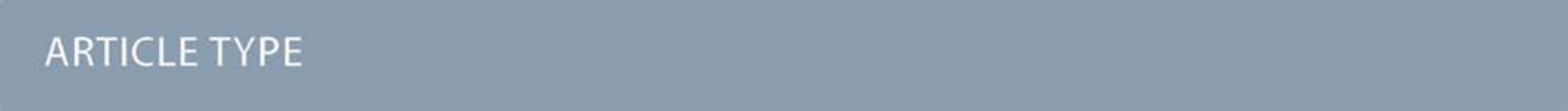}}
\fancyhead[L]{\hspace{0cm}\vspace{1.5cm}\includegraphics[height=30pt]{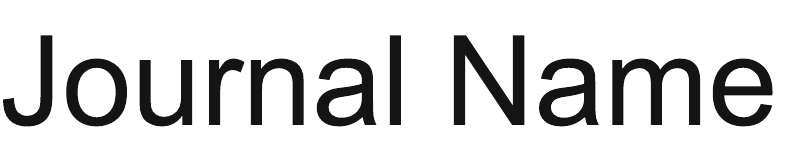}}
\fancyhead[R]{\hspace{0cm}\vspace{1.7cm}\includegraphics[height=55pt]{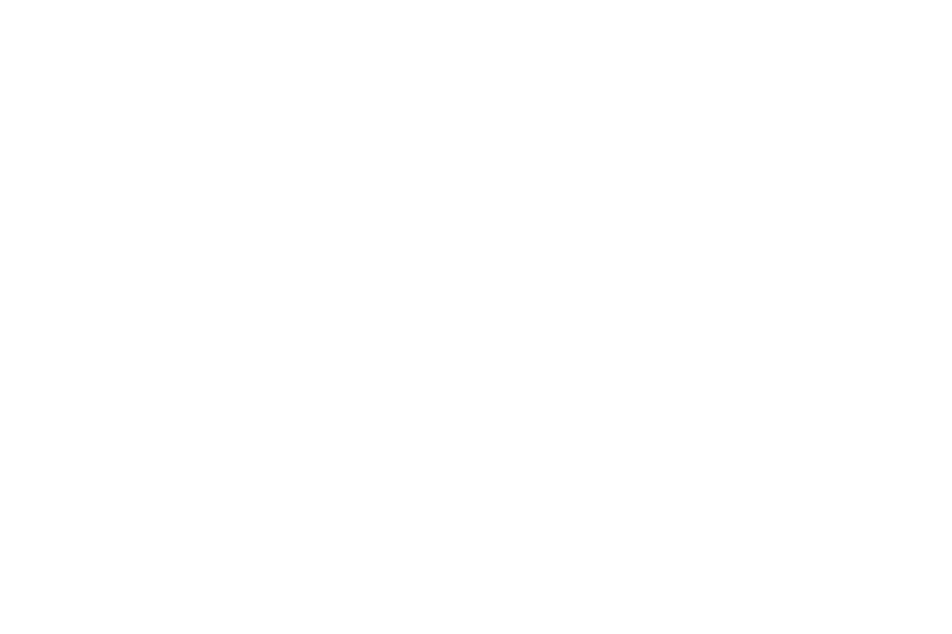}}
\renewcommand{\headrulewidth}{0pt}
}

\makeFNbottom
\makeatletter
\renewcommand\LARGE{\@setfontsize\LARGE{15pt}{17}}
\renewcommand\Large{\@setfontsize\Large{12pt}{14}}
\renewcommand\large{\@setfontsize\large{10pt}{12}}
\renewcommand\footnotesize{\@setfontsize\footnotesize{7pt}{10}}
\makeatother

\renewcommand{\thefootnote}{\fnsymbol{footnote}}
\renewcommand\footnoterule{\vspace*{1pt}%
\color{cream}\hrule width 3.5in height 0.4pt \color{black}\vspace*{5pt}} 
\setcounter{secnumdepth}{5}

\makeatletter 
\renewcommand\@biblabel[1]{#1}            
\renewcommand\@makefntext[1]%
{\noindent\makebox[0pt][r]{\@thefnmark\,}#1}
\makeatother 
\renewcommand{\figurename}{\small{Fig.}~}
\sectionfont{\sffamily\Large}
\subsectionfont{\normalsize}
\subsubsectionfont{\bf}
\setstretch{1.125} 
\setlength{\skip\footins}{0.8cm}
\setlength{\footnotesep}{0.25cm}
\setlength{\jot}{10pt}
\titlespacing*{\section}{0pt}{4pt}{4pt}
\titlespacing*{\subsection}{0pt}{15pt}{1pt}

\fancyfoot{}
\fancyfoot[LO,RE]{\vspace{-7.1pt}\includegraphics[height=9pt]{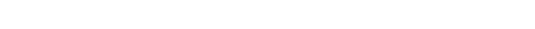}}
\fancyfoot[CO]{\vspace{-7.1pt}\hspace{13.2cm}\includegraphics{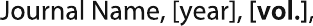}}
\fancyfoot[CE]{\vspace{-7.2pt}\hspace{-14.2cm}\includegraphics{RF}}
\fancyfoot[RO]{\footnotesize{\sffamily{1--\pageref{LastPage} ~\textbar  \hspace{2pt}\thepage}}}
\fancyfoot[LE]{\footnotesize{\sffamily{\thepage~\textbar\hspace{3.45cm} 1--\pageref{LastPage}}}}
\fancyhead{}
\renewcommand{\headrulewidth}{0pt} 
\renewcommand{\footrulewidth}{0pt}
\setlength{\arrayrulewidth}{1pt}
\setlength{\columnsep}{6.5mm}
\setlength\bibsep{1pt}

\makeatletter 
\newlength{\figrulesep} 
\setlength{\figrulesep}{0.5\textfloatsep} 

\newcommand{\topfigrule}{\vspace*{-1pt}%
\noindent{\color{cream}\rule[-\figrulesep]{\columnwidth}{1.5pt}} }

\newcommand{\botfigrule}{\vspace*{-2pt}%
\noindent{\color{cream}\rule[\figrulesep]{\columnwidth}{1.5pt}} }

\newcommand{\dblfigrule}{\vspace*{-1pt}%
\noindent{\color{cream}\rule[-\figrulesep]{\textwidth}{1.5pt}} }

\makeatother

\twocolumn[
  \begin{@twocolumnfalse}
\vspace{3cm}
\sffamily
\begin{tabular}{m{4.5cm} p{13.5cm} }

\includegraphics{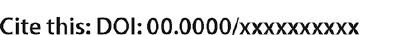} & \noindent\LARGE{\textbf{Dynamics and Clogging of colloidal monolayers magnetically driven through an heterogeneous landscape$^\dag$}} \\
\vspace{0.3cm} & \vspace{0.3cm} \\

 & \noindent\large{Sergio Granados Leyva,\textit{$^{a}$} Ralph Lukas Stoop,\textit{$^{a}$} Pietro Tierno,\textit{$^{a,b,c}$} and Ignacio Pagonabarraga$^{\ast}$\textit{$^{a,b,d}$}} \\

\includegraphics{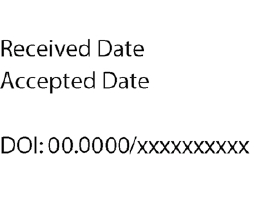} & \noindent\normalsize{We combine experiments and numerical simulations to investigate the emergence of clogging in a system of interacting paramagnetic colloidal particles driven against a disordered landscape of larger obstacles. We consider a single aperture in a landscape of immobile silica particles which are irreversibly attached to the substrate. 
We use an external rotating magnetic field to 
generate a traveling wave potential which drives 
the magnetic particles against these obstacles at 
a constant and frequency tunable speed. 
Experimentally we find that the particles display an intermittent dynamics with power law distributions at high frequencies. We reproduce these results by using numerical simulations and shows that clogging in our system arises at large frequency, when the particles desynchronize with the moving landscape. 
Further, we use the model to explore the hidden role of flexibility in the obstacle displacements and the effect of hydrodynamic interactions between the particles. We also consider numerically the situation of a straight wall and investigate the range of parameters where clogging emerges in such case.
Our work provides a soft matter test-bed system to
investigate the effect of clogging in driven microscale matter.} \\

\end{tabular}

 \end{@twocolumnfalse} \vspace{0.6cm}

  ]

\renewcommand*\rmdefault{bch}\normalfont\upshape
\rmfamily
\section*{}
\vspace{-1cm}


\footnotetext{\textit{$^{a}$~Departament de F\'isica de la Mat\`eria Condensada, Universitat de Barcelona, Av. Diagonal 647, 08028, Barcelona, Spain. Tel: +34 934034031; E-mail: ptierno@ub.edu}}
\footnotetext{\textit{$^{b}$~Universitat de Barcelona Institute of Complex Systems (UBICS), Universitat de Barcelona, Barcelona, Spain.}}
\footnotetext{\textit{$^{c}$~Institut de Nanoci$\grave{e}$ncia i Nanotecnologia, IN$^2$UB, Universitat de Barcelona, Av. Diagonal 647, 08028, Barcelona, Spain.}}
\footnotetext{\textit{$^{d}$~Centre Europ\'een de Calcul Atomique et Mol\'eculaire, \'Ecole Polytechnique F\'ed\'erale de
Lausanne (EPFL), 1015 Lausanne, Switzerland.}}

\footnotetext{\dag~Electronic Supplementary Information (ESI) available: two videoclips, one from the experiments and the other from numerical simulation 
illustrating the dynamics of the driven magnetic colloids. See DOI: 00.0000/00000000.}


\section*{Introduction}
Understanding the transport properties of 
microscopic particles trough 
heterogeneous media~\cite{Isichenko1992,Snarskii2016} is important for several 
technological processes, including filtration,~\cite{Redner2000,Roussel2007}
particle sorting,~\cite{MacDonald2003,Martinez2016} microfluidics~\cite{Squires2005,Whitesides2006} and  
many others across material and engineering science.  
From a fundamental point of view, the are several
fascinating nonequilibrium 
phenomena that emerge when such particles 
are driven across disordered landscapes such as depinning, jamming, plastic flow and rectification effects.~\cite{Reichhardt2017}
These phenomena are also common to other 
physical systems across different length scales, from 
vortex matter driven across type II superconductors,~\cite{Cohen1997} to electrons on liquid helium,~\cite{Rees2012} active matter~\cite{Bechinger2016,Parisi2018} and skyrmions.~\cite{Reichhardt2015}
The simple case of an ensemble of particles forced to pass through a single constriction may give rise to different complex effects, such as intermittency in particle flow, 
clogging and complete blockage via formation of arches and particle bridges.~\cite{Loz12} 
This effect has been investigated in the past 
on different length scales,~\cite{Zuriguel2015} from humans,~\cite{Helbing1995} to sheeps,~\cite{Garcimartin2015} granular particles~\cite{Jan09,To01}  and biological systems.~\cite{Galajda8704,Koumakis2013,Reichhardt2018}
While most of the work has focused on clogging in 
macroscopic systems, only few studies have addressed the case of microscopic particles dispersed in fluid media.~\cite{Haw2004,Zuriguel2015,Marin2018}
At such scale the presence of thermal fluctuation and hydrodynamic interactions between the particles may alter the system dynamics reducing 
or favoring the effect of clogging.  

In this context, some of us~\cite{Stoop18} investigated recently the collective dynamics of paramagnetic colloidal particles that were driven across a disordered landscape of obstacles, namely non magnetic particles fixed at random positions above a surface.  
While this work focused on the global transport properties 
in the presence of several obstacles, it did not include the detailed study of a single aperture, where the particle flow is not perturbed by the presence of several openings. Moreover, investigating a single aperture could provide many insight toward understanding the occurrence of clogging in microscale matter, and has a direct connection to the other systems previously mentioned.
Thus, in this work, we
investigate clogging of colloidal
particles when they are forced to pass through a single, narrow opening.
Starting from our experimental system, we demonstrate the occurrence
of clogging, characterized
by a power-law decay of the statistical distribution of the
passage time of the particles. We complement our experimental findings
with Brownian dynamic simulations which unveil 
relevant mechanisms that determine the anomalous flow of the forced
colloids, in particular analyzing the role of hydrodynamic interactions and the flexibility of the obstacles
at the opening.
Finally, we extend the numerical simulations to consider also the situation of a narrow opening in a planar wall of non magnetic particles. While such case is difficult to realize experimentally via  direct particle sedimentation, it is similar to many other situations 
present in
e.g. microfluidics systems, which are characterized by fixed and straight PDMS channels. 

\section*{Methods}
\subsection*{Experimental system}
%
\begin{figure}[t]
\begin{center}
\includegraphics[width=\columnwidth,keepaspectratio]{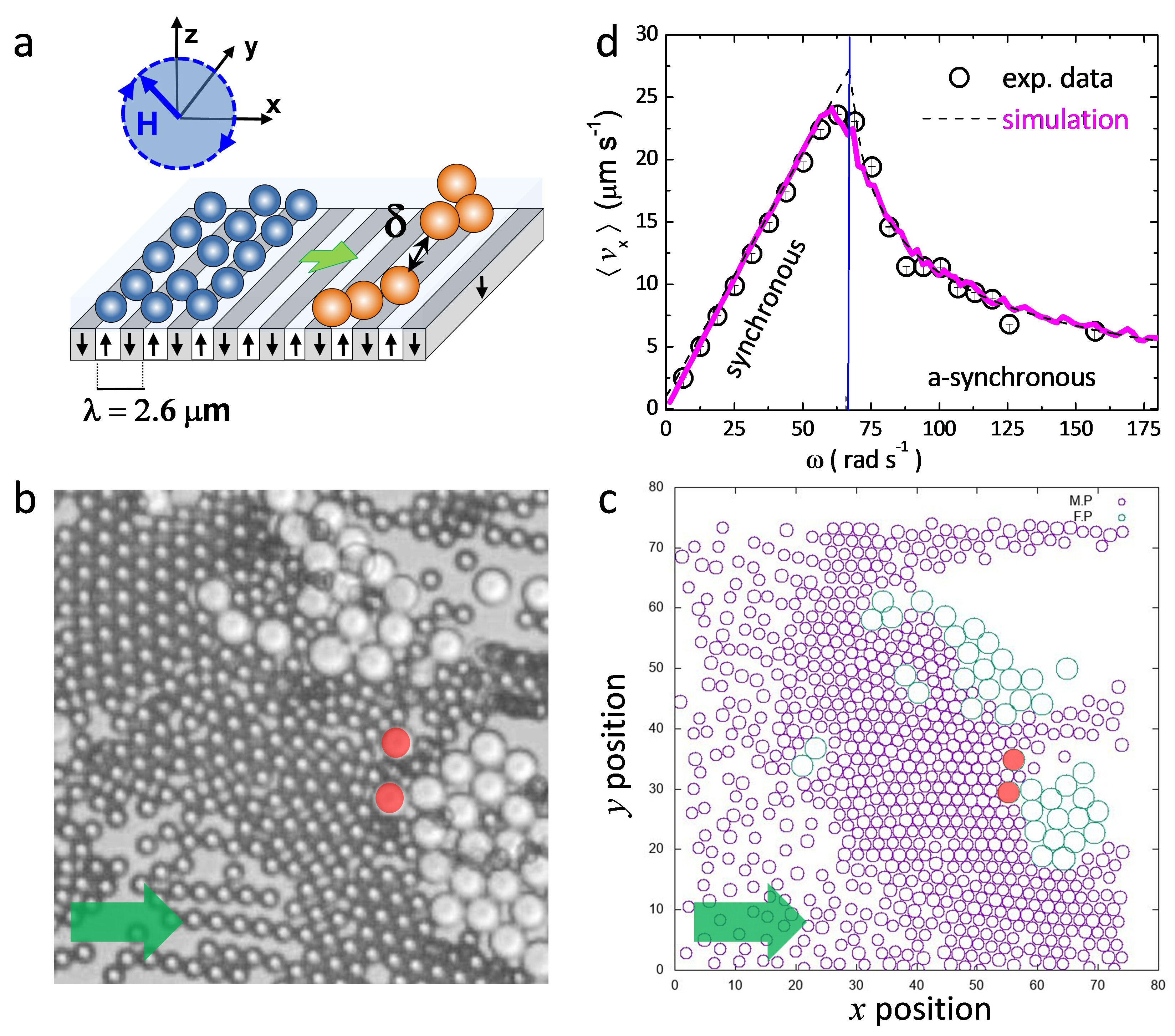}
\caption{(a) Schematic showing the experimental system composed of a monolayer of paramagnetic colloids (blue)  driven against larger silica particles (orange) and arranged to form one opening of width $\delta$.
The particles are located above a ferrite garnet film of wavelength $\lambda$ and are driven toward right by a rotating magnetic field elliptically polarized in the 
$(\hat{x},\hat{z})$ plane. 
(b,c) Experimental (b) and simulation (c) images showing a portion of the whole system where the silica particles (larger colloids) form 
a constriction of width $\delta$. The direction of motion is indicated by a green arrow. See VideoS1 and Video S2 in the Supporting Informations.
(d) Mean particle speed 
$\langle v_x \rangle$ along the direction of motion versus angular frequency 
$\omega$ in 
the absence of obstacles. Scattered circles are experimental data, while continuous magenta line are results from numerical simulation. The continuous blue line at 
$\omega_c= 68.8$ rad s$^{-1}$ separates the synchronous (left) 
and the asynchronous (right) regimes. Dashed lines are fit following the equations in the text.}
\label{figure1}
\end{center}
\end{figure}
We use commercially available paramagnetic colloidal particles (Dynabeads, M-270, Dynal) characterized by a 
diameter of $d=2.8\mu$m and  a magnetic volume susceptibility of 
$\chi \sim 0.4$. The particles are composed of a polystyrene matrix with surface carboxylic groups, and 
are doped with nanoscale iron oxide grains.
The particles are dispersed in highly deionized water (MilliQ, Millipore) and deposited above the surface of 
a uniaxial ferrite garnet film (FGF). The FGF was previously synthesized via dipping liquid phase epithaxy~\cite{Tierno2009} and 
it is characterized by parallel stripes of ferromagnetic domains with alternating up and down magnetization.
In the absence of an external field, the wavelength of the FGF pattern is $\lambda=2.6\rm{\mu m}$, and the saturation magnetization $M_s=1.3 \cdot10^4\rm{Am^{-1}}$, see Fig.~\ref{figure1}(a). 
Before the experiments, the FGF is coated with a 
thin layer ($\sim1\rm{\mu m}$ thick) of a photoresist (AZ-1512 Microchem, Newton, MA) using spin coating and backing procedures.~\cite{Tierno2012} 

The particles are transported against fixed obstacles 
made of silica dioxide microspheres (44054-5ML-F, Sigma-Aldrich) with diameter $d=5\rm{\mu m}$ (standard deviation 
$\sigma \lesssim 0.35 \rm{\mu m}$). Before the experiments, the silica particles 
are irreversibly attached above the FGF surface with the following procedure. 
First, the silica particles are diluted in highly deionized water at different concentrations and deposited above the FGF surface. After their sedimentation which lasts few minutes, the particles float 
above the FGF due to their negative surface charge.
Sticking on the substrate is induced by the addition of a salt, namely  a solution of $10$mM NaCl in water. The NaCl ions of the salt screen the electrostatic interactions favoring permanent linkage of the silica particles to the FGF via attractive van der Waals interactions. After that, the salt solution was removed and was replaced by a water dispersion containing the paramagnetic colloids. 
As a result, the magnetic particles float on a substrate composed of a quenched disorder of silica obstacles, Fig.~\ref{figure1}(b). 

\subsection*{Transport mechanism}

We start by describing the particle motion in the absence of obstacles.
Once placed above the FGF surface, the particles are attracted by the magnetic domain walls, and form a two-dimensional monolayer above the $(\hat{x},\hat{y})$ plane. We induce directed particle transport by using an external rotating magnetic field elliptically polarized in a perpendicular plane  $(\hat{x},\hat{z})$, see Fig.~\ref{figure1}(a). 
The applied field, with amplitude $H_0$ and angular frequency $\omega$ is given by,
$\bm{H}\equiv (H_x \cos{(\omega t)}\hat{x},0,H_z \sin{(\omega t)}\hat{z})$,
where $H_0=\sqrt{(H_x^2+H_z^2)/2}$, and $(H_x,H_z)$ are the two field components. Since the field is elliptically polarized, it can be characterized also by the ellipticity $\beta=(H_x^2-H_z^2)/(H_x^2+H_z^2)$,
where $\beta \in [-1,1]$ and $\beta=0$ corresponds to the circular polarization. For all the experiments we 
keep fixed the amplitude to $H_0=800\rm{A m^{-1}}$ 
and the ellipticity parameter to $\beta=-0.4$. The latter choice ensures  
that dipolar interactions are negligible along the propulsion direction~\cite{Str14}
and the particles can be considered, in first approximation, as 
hard-spheres.
 
The external magnetic field modulates the stray field of the FGF surface, 
and it generates a two-dimensional sinusoidal-like 
potential which continuously translates along one direction ($\hat{x}$) perpendicular to the magnetic stripes.
The potential moves at a constant and frequency tunable speed $v_p= \omega \lambda/(2 \pi)$, and drags the paramagnetic colloids located in its minima with it. 
As shown in Fig.~\ref{figure1}(d), depending on the driving frequency one 
can identify two dynamic regimes.~\cite{Str13} Below a critical frequency $\omega_c$, the 
particles are trapped in the potential minima and move with these minima (synchronous regime) at a constant average speed $\langle v_x \rangle=v_p$. When $\omega >\omega_c$,  
the motion of the particles desynchronizes with the moving landscape, and the average speed decreases as $\langle v_x \rangle=v_p (1-\sqrt{1-(\omega_c/\omega)^2})$ 
(asynchronous regime). As shown in Fig.~\ref{figure1}(d), we find $\omega_c= 68.8 \rm{rad \, s^{-1}}$ for the experimental parameters used here, a value which is in excellent agreement with the numerical simulations (see later) for an obstacle free system. 
Further, in the synchronous regime the particles acquire a translational speed $\langle v_x \rangle \in [2,23] \rm{\mu m s^{-1}}$ which corresponds to a P\'eclet number $Pe \in [20,230]$. We estimate the latter by considering the ratio between the Brownian time $\tau_B=d^2/(4 D_{eff})$ required by the particle to diffuse its own radius ($d/2$), and the driven time $\tau_D=d/(2 \langle v_x \rangle)$ required to move its radius due to the magnetic landscape. Here we use the value of the effective diffusion coefficient $D_{eff}=0.14 \rm{\mu m ^2 s^{-1}}$ which was estimated in a previous work~\cite{Tierno2007}.

\subsection*{Simulation scheme}
We complement the experimental results by using
Brownian dynamic simulation with periodic boundary conditions. We consider a system composed of moving paramagnetic colloids, of size  $r_p$,  and silica obstacles, of size $r_o$.
All particles are characterized by their positions, $\bm{r}_i$ with $i=1...N_p+N_o$. Initially, we consider obstacles fixed on the substrate,  $\bm{r}_i=\bm{r}_{i,eq}$, $i=N_p,...,N_p+N_o$, while the magnetic particles evolve following the overdamped dynamics:
\begin{equation}
\frac{1}{\mu} \frac{d\bm{r}_i}{dt}=\sum_j \bm{F}^{int}(\bm{r}_{ij})+\bm{F}^{ext}(\bm{r}_{i})+\bm{F}^{T}(\bm{r}_{i}) \, \, \, ,
\label{motion}
\end{equation}
where $i=1,...,N_p$,
$\mu$ is the particle mobility, $\bm{F}^{int}$
is the pair interaction between the colloids, $\bm{F}^{ext}$ the external driving force and $\bm{F}^T$ accounts for the force exerted by the  thermal bath.
The interparticle forces derive from a Yukawa potential, and account effectively  for the colloidal electrostatic short range repulsion and finite particle size. The force between a particle $i$ of type $a$, of radius $r_a$, and a particle $j$ of type $b$, of radius $r_b$, can be written as:
\begin{equation}
{\bf{F}}^{int}(r_{ij})=\frac{U_{ab}}{\lambda_{ab}}\sum_{i\neq{j}}^N\left[\frac{\sigma_{ab}}{r_{ij}}\left( \frac{\sigma_{ab}}{r_{ij}} +\frac{\sigma_{ab}}{\lambda{_{ab}}}e^{-\frac{r_{ij}}{\lambda_{{ab}}}}\right)-B_{ab}\right]{\bf{e}_{r_{ij}}} \, \, ,
\label{force}
\end{equation}
where ${\bf{e}_{r_{ij}}}$ a unit vector along the two considered particles, and $\bm{r}_{ij}=\bm{r}_i-\bm{r}_j$.
The parameter $U_{ab}$ quantifies the strength, and $\lambda_{ab}$   the characteristic decay length,   of the Yukawa potential between the interaction of a particle of type $a$ and one of type $b$, $\sigma_{ab}=(r_a+r_b)/2$ denotes the radius of the interaction between the pair. The parameter $B_{ab}$ is a constant ensuring that the force is zero at the cutoff interaction radius $r_c$, 
\begin{equation}
B_{ab}=\frac{\sigma_{ab}}{r_c}e^{-\frac{r_c}{\lambda_{ab}}}(\frac{\sigma_{ab}}{\lambda_{ab}}+\frac{\sigma_{ab}}{r_c}) \, \, .
\end{equation}
We neglect the magnetic dipolar interactions since it was previously shown~\cite{Str14} that these interactions become negligibly small along the propulsion direction for applied fields with $\beta=-1/3$, similar to our case. 
The  external force results from the magnetically modulated landscape and is given by~\cite{Str14}
\begin{equation}
{\bf F}^{ext}(x,t)=F_M\left[u_1(t)\sin\left(\frac{2\pi{x}}{\lambda}\right)-u_2(t)\cos\left(\frac{2\pi{x}}{\lambda}\right)\right]\bf{\hat{x}}
\label{ext_pot}
\end{equation}
where $u_1(\beta,t)=\sqrt{1+\beta}\cos(\omega t)$,  $u_2(\beta,t)=\sqrt{1-\beta}\cos(\omega t)$ and 
$F_M = 16 H_0 e^{-2\pi z/\lambda}/(\lambda M_s)$
is a prefactor that considers the particle elevation ($z$) from the substrate. 
The last term in Eq.~\ref{motion} is a random force associated to the temperature. Integration of this term over one time step, $\Delta t$, gives a random displacement $\Delta{\bf{r}}^r$ characterized by a Gaussian distribution, of magnitude $\Delta{\bf{r}}^r=\mu{\bf{F}}_i^T\Delta t$ and $<(\Delta r^r)^2>=4 D_0 \Delta t$, being $D_0$ the particle diffusion coefficient.

For computational convenience, we make Equation~\ref{motion}
adimensional using the magnitude of the external force $F_M$, the colloid radius $r_p$, the characteristic speed
$v_c=F_M\mu$ and time $\tau_D=d/2\mu{F}_M$ scales, which identify the relevant P\'eclet number, $Pe= v_{c} r_p/D_0={\mu}F_M\sigma/D_0$. Accordingly, we can  rewrite  Eq.~\ref{motion} as:
\begin{equation}
\frac{d\bar{\bm{r}_i}}{d t}=\frac{U_{ab}}{\lambda_{ab}F_M}\sum_{j\neq i}{\bf{\bar{{\pmb{\psi}}}}}(\bar{\bm{r}}_{ij})+
{\bf{\bar{F}}}^{ext}(\bar{x},\bar{t}) \\ +\sqrt{\frac{4}{Pe}\frac{\tau_D}{\Delta{t}}}\pmb{\bar{{\xi}}}
\label{eq_red_unit}
\end{equation}
where $\bar{\bm{r}}=\bm{r}/r_p$,  $\bar{\omega}=\omega \tau_D$, $\bar{\Delta t}=\Delta t/\tau_p$.,  In turn, ${\bf{\bar{{\pmb{\psi}}}}}={\bf{\bar{{\pmb{\psi}}}}}(\bar{\bm{r}})_{ij}$  is the dimensionless  Yukawa force between particles $i$ and $j$,
\begin{equation}
{\bf{\bar{{\pmb{\psi}}}}}(\bar{\bm{r}}_{ij}) = \left[\frac{\bar{\sigma}_{ab}}{\bar{r}_{ij}}\left( \frac{\bar{\sigma}_{ab}}{\bar{r}_{ij}} +\frac{\bar{\sigma}_{ab}}{\bar{\lambda}_{ab}}e^{-\frac{\bar{r}_{ij}}{\bar{\lambda}_{{ab}}}}\right)-B_{ab}\right]{\bf{e}_{r_{ij}}}
\end{equation}
where $\bar{\sigma}_{ab}=\sigma_{ab}/r_p$, and  $\bar{\lambda}_{ab}=\lambda_{ab}/r_p$. Finally,  ${\bf{\bar{F}}}^{ext}(\bar{x},\bar{t})$ corresponds to the dimensionless form of the external, driving force, Eq.~\ref{ext_pot}.

We integrate Eq.~\ref{eq_red_unit} using a time step 
$\Delta t/\tau_D = 1\cdot 10^{-4}$,  $r_p=1$, $r_o=1.78$,
$\lambda_{p}=1$, $\lambda_{op}=1.39$, $U_{pp}/\lambda_{pp}F_M=300$, $U_{op}/\lambda_{op}F_M=150$  and Pe$=150$. 
Further parameters were extracted by fitting 
the results from the experimental data to the one obtained
from numerical simulation of an obstacle free system, see Fig.~\ref{figure1}(d). 
From this benchmarking, we estimate 
the mobility $\mu\simeq{1}/6\pi\eta{r_a}$ of the paramagnetic particles in water as $\mu=3.79\cdot10^{7} \rm{N \, s \, m^{-1}}$  and the magnitude of the force
of the traveling wave potential as $F_M= 0.1$pN . With these values we calculate the characteristic time of the particle motion as $\tau_D=0.04$s. 
Having
demonstrated a good agreement between the simulations and the experiments
in the absence of obstacles, we introduce the obstacles to our simulations. As shown in Fig.~\ref{figure1}(c),  we map the experimental
situation (Fig.~\ref{figure1}(b)) to our simulations
by using the same spatial distribution of the obstacles.

In the numerical simulations obstacles are considered as rigidly attached to  the substrate, and $N$ then runs only over paramagnetic colloids. Careful inspection of the experimental videos reveals very small oscillations of the silica particles around their equilibrium position upon collisions. These oscillations may result from a combination between steric interactions of the silica particles with the polymer coated film and weak Van der Waals attractions. Quantifying these interactions is a difficult task, and we consider them as an effective spring constant which accounts for the fact that the silica particles are not pulled away when subjected to forces coming from the driven colloids. Moreover, these oscillations are very small, and their amplitude is smaller than the error bars associated to the tracking. To account for this effect, we   take into account the possibility that the obstacles are slightly displaced due to the forces exerted by the moving paramagnetic colloids. In  this case we include the obstacles as part of the $N$ moving particles, which evolve according to the equation:
\begin{equation}
\frac{d \bar{\bm{r}}_i }{dt}=\sum_{j\neq i}\frac{U_{ab}}{\lambda_{ab}F_M\mu}{\bf{\bar{{\pmb{\psi}}}}}(\bar{\bm{r}_{ij}})+\frac{k r_p}{\mu{F_M}}({\bf{\bar{r}}}_i-{{ \bf \bar{r}}_{i,eq}}) \, \, \, ,
\end{equation}
with  $i=N_p+1,...,N_p+N_o$.
The particles are subject to a spring of strength $k$ whenever the obstacle is displaced from its equilibrium position, ${\bf \bar{r}}_{i,eq}$. The competition between the external driving frequency and the one related with the elastically displaced particles can lead to different and significant results.

Since the colloids are embedded in a solvent, we have also generalized the Brownian dynamics of the paramagnetic colloids to accounts for the impact that hydrodynamic interactions mediated by the solvent has in the  clogging kinetics. Each particle is dragged by the fluid flow, $\bm{v}_i^H$, generated by the rest of  the  colloids due to the forces they are  subject to, an effect that can be captured generalizing eq.(\ref{motion}) to 
\begin{equation}
\frac{1}{\mu} \frac{d\bm{r}_i}{dt}=\sum_j \bm{F}^{int}(\bm{r}_{ij})+\bm{F}^{ext}(\bm{r}_{i})+\bm{F}^{T}(\bm{r}_{i})+\frac{\bm{v}_i^H}{\mu} \, \, \, ,
\label{motion_HI}
\end{equation}
where  we take into account the  impact that the close proximity of the wall, and express the induced velocity in the point particle approximation as:
\begin{equation}
\bm{v}_i^H=\sum_{j\neq{i}}^N{{\bf{G}}({\bf{r}}_{ij})}{\bf{F}}^{int}(\bm{r}_{ij}) \, \, \, .
\label{motion_Blake}
\end{equation}
Here where ${\bf{G}}({\bf{r}}_{ij})$ stands for the Blake-Oseen mobility tensor~\cite{Blake} which considers the effect of the close proximity of the substrate in a far field approximation.
We note that a sample with finite size could produce a screening of the hydrodynamic interactions between the particles. However, in our case the experimental cell is wide enough that this screening develops over length scales that are long compared with the distances over which the particles interact and move through the obstacles.

\begin{figure}[t]
\begin{center}
\includegraphics[width=\columnwidth,keepaspectratio]{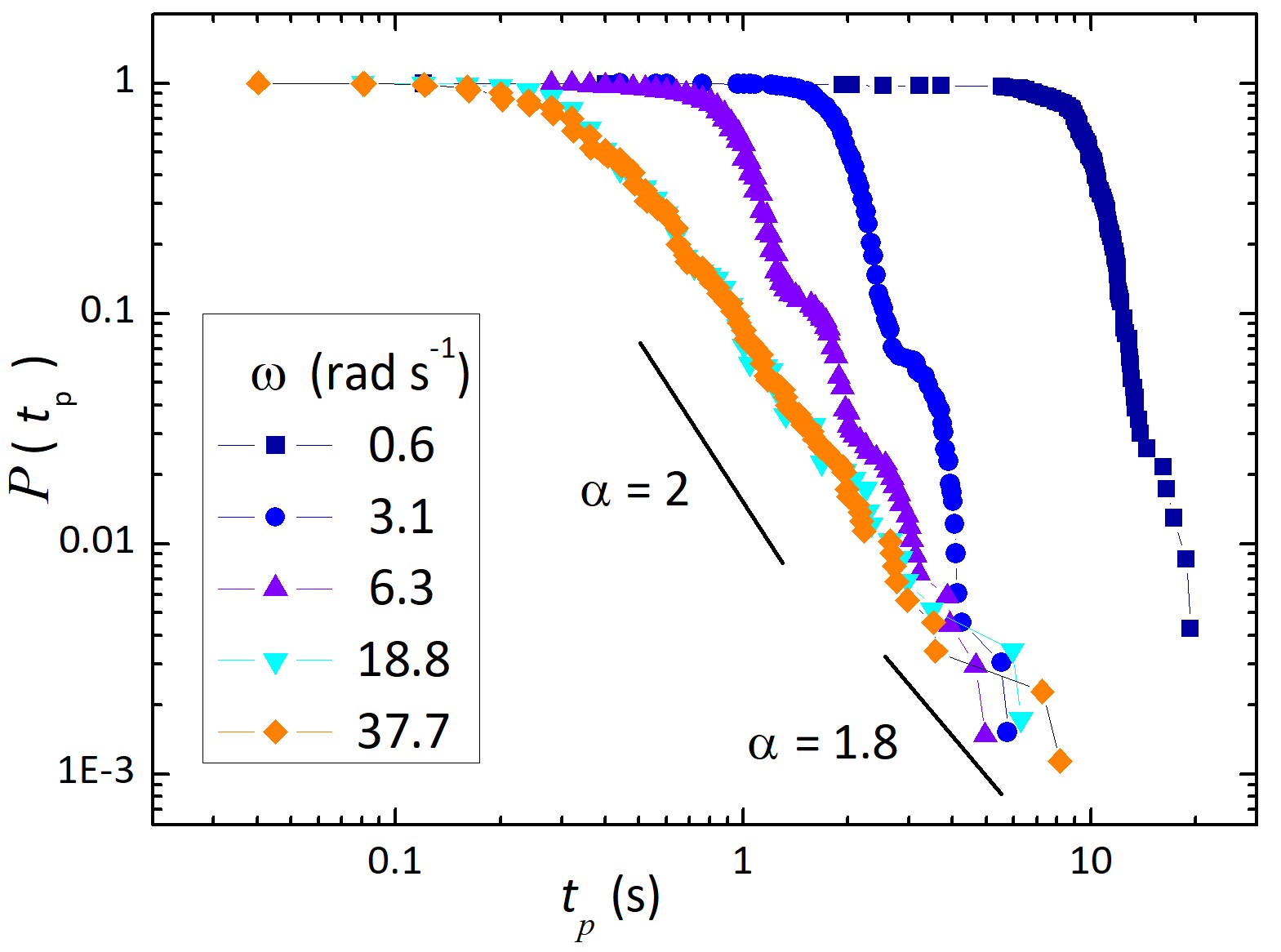}
\caption{Experiments: The distribution function $P(t_p)$ 
of time lapses $t>t_p$  for magnetic particles passing through a single constriction and at different driving frequencies. The continuous black line indicates the slope $\alpha = 2$
of the power-law $P\sim t_p^{-\alpha}$ which is used to distinguish clogged state ($\alpha < 2$)
and unclogged ones  ($\alpha > 2$).}
\label{figure2}
\end{center}
\end{figure}

\section*{Discussion}

\subsection*{Clogging: experimental results}
Introducing obstacles to the substrate
completely changes the behavior of the driven particles as they
are now forced to interact with the silica spheres which induces a
deviation of their trajectories.
As shown in Figs.~\ref{figure1}(c,d), we consider the case of obstacles which create a narrow gap of width $\delta$.
We find that the presence of the opening strongly reduces the average speed of the monolayer 
eventually even leading to complete blockage, $\langle v_x \rangle =0$, for 
$\delta<3 \rm{\mu m}$.
For larger values of the distance $3\rm{\mu m} <\delta< 4 \rm{\mu m}$,
we find an intermittent flow 
of the magnetic colloids 
which arises  from the simultaneous arrival
of the particles  
at the aperture, 
and their accumulation 
in a close packed state, which is locally jammed.

The appearance of clogging, namely the blockage of the 
particle flow, can be characterized by measuring 
the distribution $P(t_p)$ that quantifies when the time lapse between the passage of consecutive 
particles through the aperture is larger than a given time $t_p$. This distribution is also known as the complementary cumulative distribution function.
It has been previously shown that in the presence of an intermittent flow 
such distribution is expected to display asymptotically
an universal behavior, i.e. a power law 
decay at high values of $t_p$ ($P\sim t_p^{-\alpha}$)
as a function of its exponent, $\alpha$.~\cite{Saloma2003,Jan09} This exponent can be used to distinguish the regime of normal flow, $\alpha  > 2$, i. e. when the average flow rate is  finite,  from intermittency  and clogging, $\alpha < 2$,  i. e. when 
the average time lapse between consecutive passing elements trough the aperture diverges. Hence, we can use $\alpha=2$ to identify the transition to clogging, where it cannot be specified whether or not there is flow at a give time.~\cite{Zuriguel2015}  
Indeed a clogged state it is not fully blocked, as some material can be briefly released. 

We use video microscopy to 
precisely track the positions of the particles and to calculate 
the corresponding distributions.
Fig.~\ref{figure2} shows $P(t_p)$ measured for different values of the driving frequency $\omega$ and keeping constant the amplitude of the applied field. 
We find two different types of behaviour depending on $\omega$ which reflect the presence in the system of the underlying magnetic domain walls. For small frequencies, namely $\omega < 6.3 \rm{rad \,  s^{-1}}$, clogging events are rare, and the distribution $P$ displays an exponential decay. This effect can be understood by considering particle trajectories which are characterized by a sequence of discrete jumps between the domain walls during each driving period. These jumps emerge in the system as periodic oscillations of the particles which are superimposed to the net drift velocity. The presence of such vibrations significantly reduces the clogging probability and the eventual formation of particle bridges. Such effect can be considered as an additional AC signal superimposed to the DC drift and allows to break the formation of a close packed monolayer of particles close to the constriction.~\cite{Jan09}. We note that a similar strategy, namely the use of an additional AC field to fluidized the system, was recently introduced in numerical simulation to maximize flow in a random obstacle array.~\cite{Reichhardt2018}  In contrast, at high frequencies, the particles still move across the domain walls but their trajectories are continuous as the discrete jump are smeared out. As a consequence, the vibrations of the domain walls are not influencing the particle movement and cannot fluidize the system close to the constriction. Thus, we find that $P$ becomes a power law, $P\sim t_p^{-\alpha}$ and the systems is more prone to display clogging behavior.
For these situations, 
we determine a common exponent $\alpha = 1.8$
for the tails of the high frequency distributions.~\cite{Cla09}
Such value is slightly higher as compared to 
pedestrian or granular systems that display 
stable arches at the constriction.~\cite{To01,Loz12}
The specificity of the colloidal system, and the role that the hydrodynamic interactions may have among the relative motion of the colloids close to the constriction may explain that the observed clogged stated are more fragile than those observed on systems composed by larger constituents.
Moreover, we find that all curves above $\omega = 6.3 \rm{rad s^{-1}}$ collapse, and clogging becomes independent on the particle speed. We note that experimental limitations impeded us to have enough statistical data, and thus to precisely determine the exponent of the distribution tails. Such limitations result from the unavoidable sticking of the particles to the FGF film after relatively long recording periods.

\subsection*{Simulation results: fixed obstacles}

\begin{figure}[t]
\begin{center}
\includegraphics[width=0.9\columnwidth,keepaspectratio]{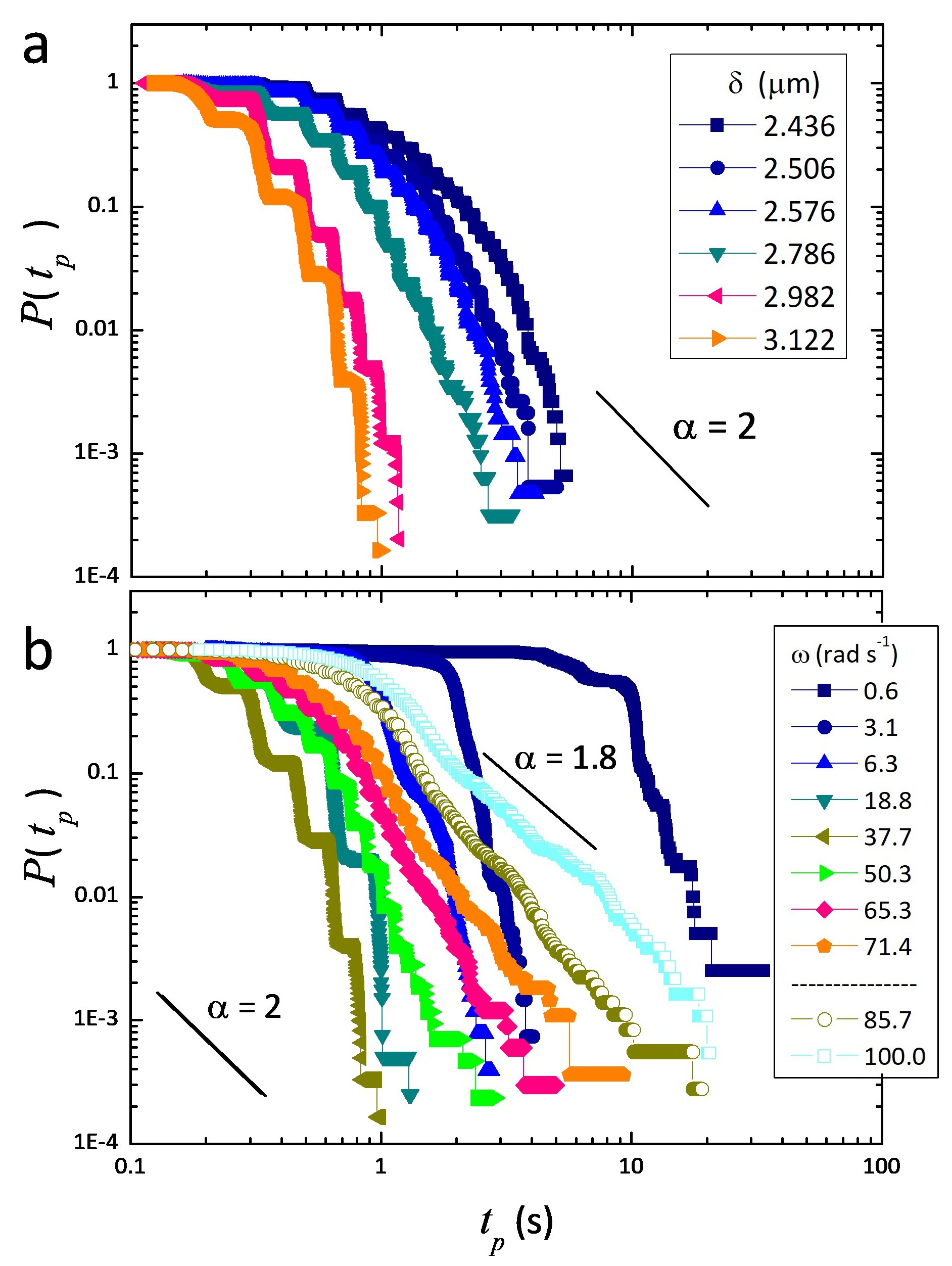}
\caption{(a,b) Numerical simulation results in the synchronous regime ($\omega = 37.7 \rm{rad \, s^{-1}}$). (a) Distribution functions $P(t_p)$ of time lapses $t_p$ for different values of the constriction width $\delta$. 
(b) Distribution functions $P(t_p)$ of time lapses $t_p$ for different values of driving frequency $\omega$, with $\delta = 2.98 \rm{\mu m}$. The results are shown for particles driven in the synchronous regime for $\omega \leq 71.4 \rm{rad s^{-1}}$, and in the  asynchronous regime for larger $\omega$.
}
\label{figure3}
\end{center}
\end{figure}

Figs.~\ref{figure3}(a,b) display the results 
from the numerical simulations for fixed  obstacles.
In Fig.~\ref{figure3}(a) we show how the distribution function 
$P(t_p)$ changes by varying the distance between the obstacles forming the constriction, at fixed driving frequency. As expected, a narrower constriction increases the particle mean passing time. Similar to the experimental data, we also observe a periodic decrease of the distributions 
followed by a series of characteristic plateaus which result from the external sinusoidal forcing. Indeed the time between two consecutive decays is proportional to the period of the applied field, $T=1/f$. These plateaus are more evident in the simulations  than in the experiments (Fig.~\ref{figure2}), due to the  higher temporal resolution achieved numerically. In the plateaus colloids have a 
higher probability to pass through the constriction, since they are located in a place where the potential is steeper, and 
thus feel a maximum positive force along the direction of motion. As the opening   associated to a pair of obstacles decreases, these plateaus become narrower and the corresponding distributions smoother. Further, we find that for all the cases considered here, the system never develops a power law decay. Thus, no clogged states are present in the synchronous regime, in contrast to the experimental case where $\alpha=1.8$ was observed for large $\omega$ but still slightly lower than $\omega_c$. 

\begin{figure}[t]
\begin{center}
\includegraphics[width=0.9\columnwidth,keepaspectratio]{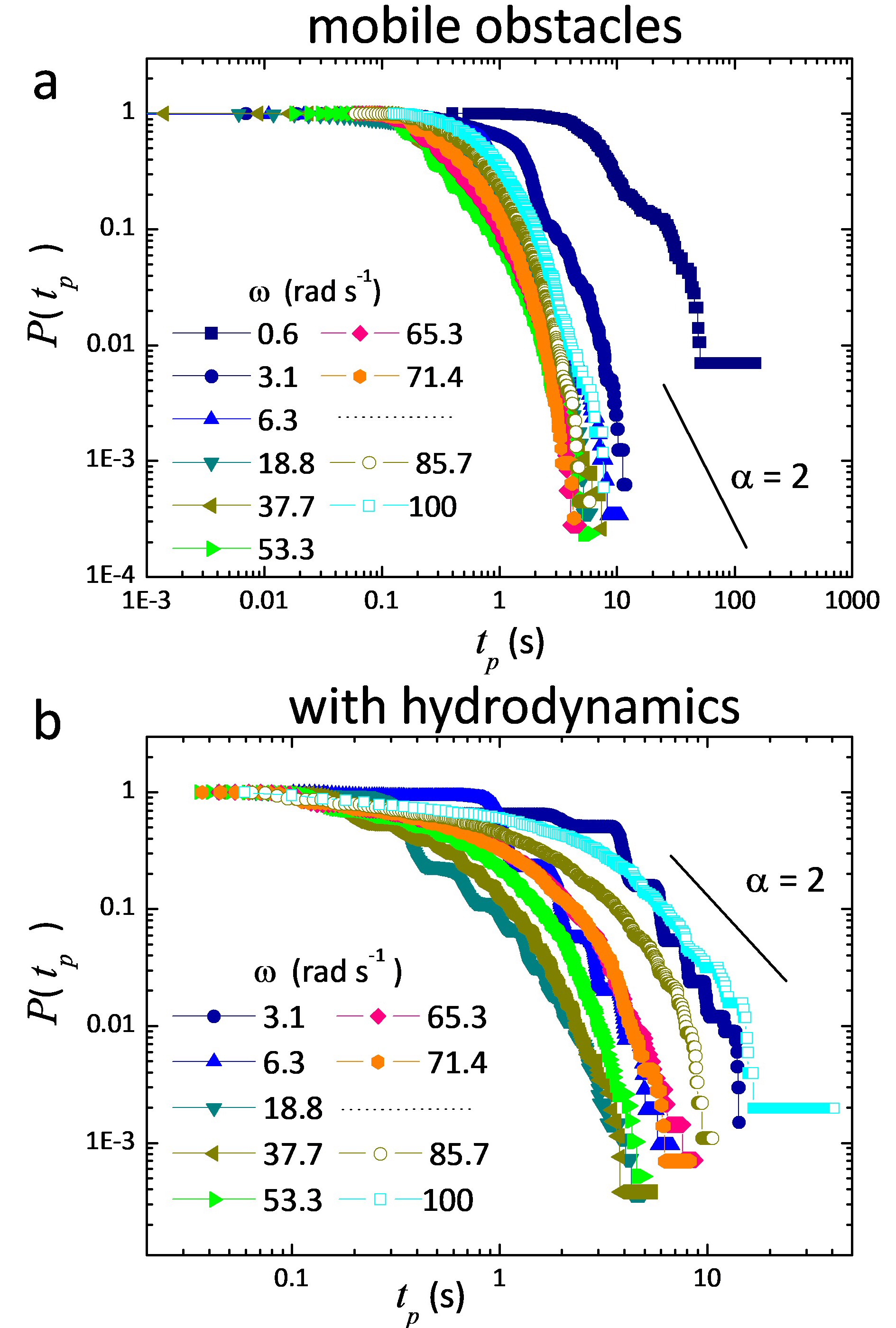}
\caption{(a) Distribution functions $P(t_p)$ of time lapses $t_p$ for different values of driving frequency $\omega$ both in the synchronous (filled symbols) and asynchronous (empty symbols) regime
from numerical simulations. The simulations were performed by considering mobile obstacles with a spring constant $k=0.593 \rm{\mu N m^{-1}}$.
The black segment indicates the slope $\alpha=2$ of the power law $P\sim t^{-\alpha}$.
(b) Distribution functions $P(t_p)$ of time lapses $t_p$ versus $\omega$  from the numerical simulations with hydrodynamic interactions.
The channel width for all data has been fixed to $\delta = 2.576 \rm{\mu m}$.}
\label{figure4}
\end{center}
\end{figure}

As shown in Fig.~\ref{figure3}(b), 
the free speed of the paramagnetic particles, proportional to the driving
frequency $\omega$ in the synchronous
regime, strongly influences the distribution of $t_p$. In fact, increasing $\omega$ shifts the time lapse distribution towards
shorter times, with a decay that may be characterized
by an exponential law similar to the experimental results in Fig.~\ref{figure2}. Interestingly, this trend  reverses
after $\omega =37.7 \rm{rad \, s^{-1}}$ as the tails of the distributions flatten and become power law but with no signature of clogging, $\alpha > 2$. Above $\omega = 85.7 \rm{rad \, s^{-1}}$,
as the particles reach the 
asynchronous region, 
the $P(t_p)$ develop power law 
tails with $\alpha \sim 1.8$  signaling clogging. This results shows that losing synchrony promotes clogging. 

\begin{figure}[t]
\begin{center}
\includegraphics[width=0.8\columnwidth,keepaspectratio]{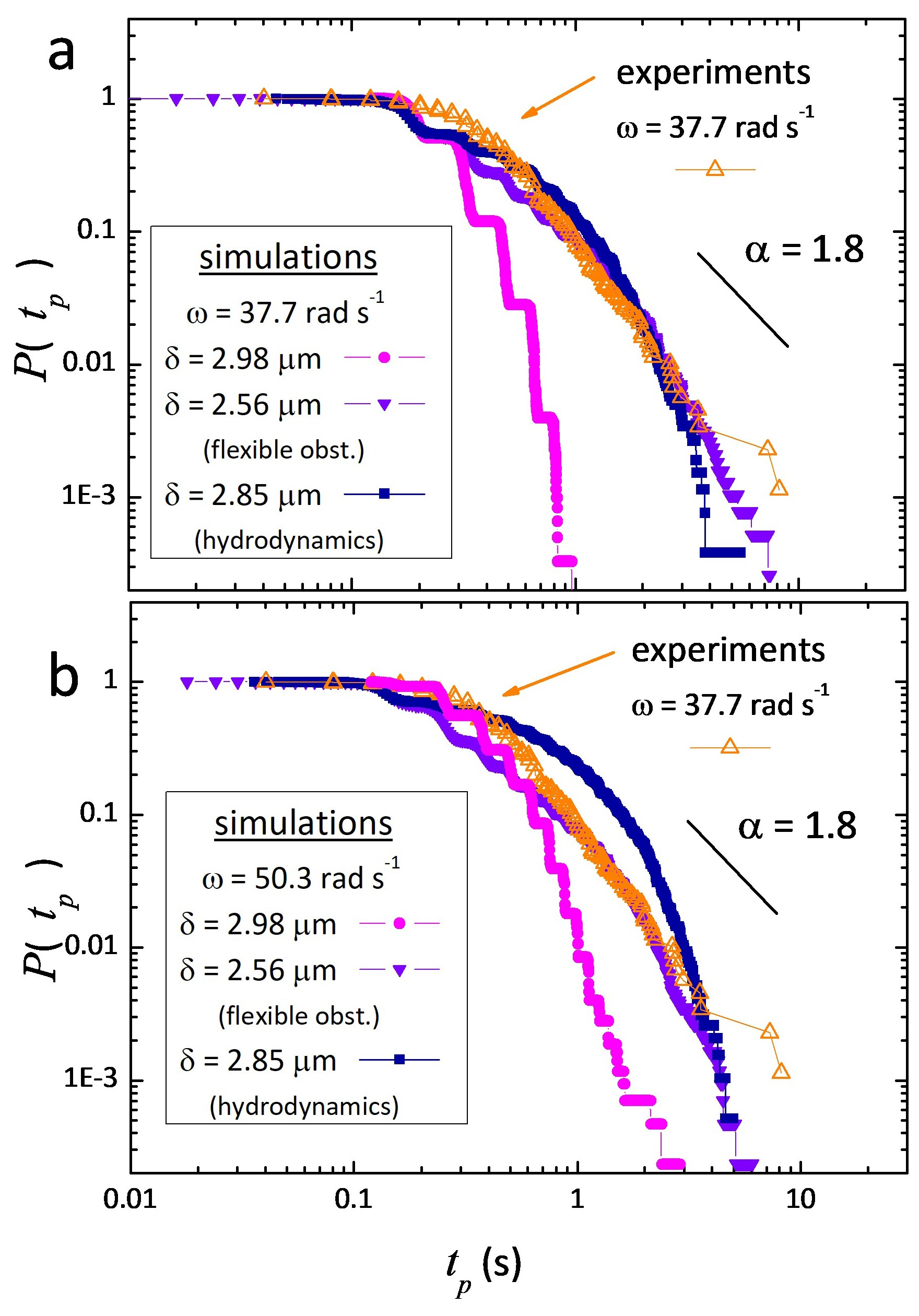}
\caption{(a,b) Comparison between the
experimental data (open symbols), frequency $\omega = 37.7 \rm{rad s^{-1}}$  and the numerical simulations (filled symbols) 
driving frequencies: (a) $\omega = 37.7 \rm{rad s^{-1}}$
and (b) $\omega = 50.3 \rm{rad s^{-1}}$.
In both images the black segment denotes the slope $\alpha = 1.8$
of the power law $P(t_p)$.}
\label{figure5}
\end{center}
\end{figure}

\subsection*{Simulation results: Harmonic spring and Hydrodynamic interactions}

The results obtained by fixing the positions of the obstacles shows the emergence of clogging at high frequencies, however they did not capture all the effects observed in the experiments as, for example, the collapse of the distributions at high frequencies 
in the synchronous regime. In order to analyze the origin of such discrepancy, we consider the impact that  moving obstacles and the hydrodynamic interactions have on the system dynamics. To this end, we first
introduce a small flexibility in the 
obstacle location and show in Fig.~\ref{figure4}(a) the corresponding results. The presence of another vibration frequency resulting from the obstacle mobility can have relevant effects and contribute directly to  the effect of clogging.
To find the optimal spring constant $k$, we have performed different simulations and compare them directly with the experimental results in order to optimize the comparison. 
We find a very similar distribution  
for a spring constant $k=0.593 \rm{\mu N \, m^{-1}}$. 
Moreover, similar to the experiments we also observe the collapse of the distributions at high frequencies by keeping fixed $k$ and varying $\omega$, Fig.~\ref{figure4}(a). 
Thus, the presence of slightly mobile
obstacles favors clogging, while using the same
constriction width with no harmonic spring gives a completely
unclogged state and a large exponent in the distribution $P(t_p)$, $\alpha>2$.
Further, as shown in Fig.~\ref{figure4}(a), we find that there are no visible plateaus as in Fig.~\ref{figure3}. 
Moreover, smaller spring constants can lead to a more clogged state than higher ones. The latter effect is unexpected, as smaller frequencies could induce longer period of time when the obstacle are more deformed and the width of the opening wider, which should favor unclogging.  
However, the results obtained in Fig.~\ref{figure4}(a) point out to an interplay between the frequency of the external magnetic field and the frequency of oscillation of the obstacles which favor clogging, e.g. obstacle  mobility leads to an interaction with other constrictions because particles passing through nearby openings, see  Fig.~\ref{figure1}(b), can also indirectly exert additional pressure on the obstacles of the opening of interest, affecting the   colloidal passage times.

Experimentally,  the paramagnetic colloids move above the FGF film generating a net flux between them and close to the constriction, see previous section for details. We use the extended Brownian Dynamics model described in the Method section to analyze the impact that  hydrodynamic interactions (HIs)  have in the cumulative colloidal passing times.
As shown in Fig.~\ref{figure4}(b), we also explore 
the effect of HIs by varying the driving frequency from the synchronous ($\omega < 85.7 \rm{rad s^{-1}}$) to the a-synchronous regime. The emergence of the finite plateaus is also observed at low frequency, while raising $\omega$
smooths the curves and produces power law tails, Fig.~\ref{figure4}(b).

The direct comparison with the experimental data taken at $\omega = 37.7 \rm{rad s^{-1}}$ is shown in Figs.~\ref{figure5}(a,b) for two frequencies both in the synchronous regime. The numerical simulations show that the introduction both of obstacle mobility and HI smooth significantly the plateaus, to the extent that they are barely visible when HI are introduced  with fixed obstacles. Moreover, 
flexible obstacles induce distribution functions which have dependence with frequency closer to the one observed experimentally.

\begin{figure}[t]
\begin{center}
\includegraphics[width=\columnwidth,keepaspectratio]{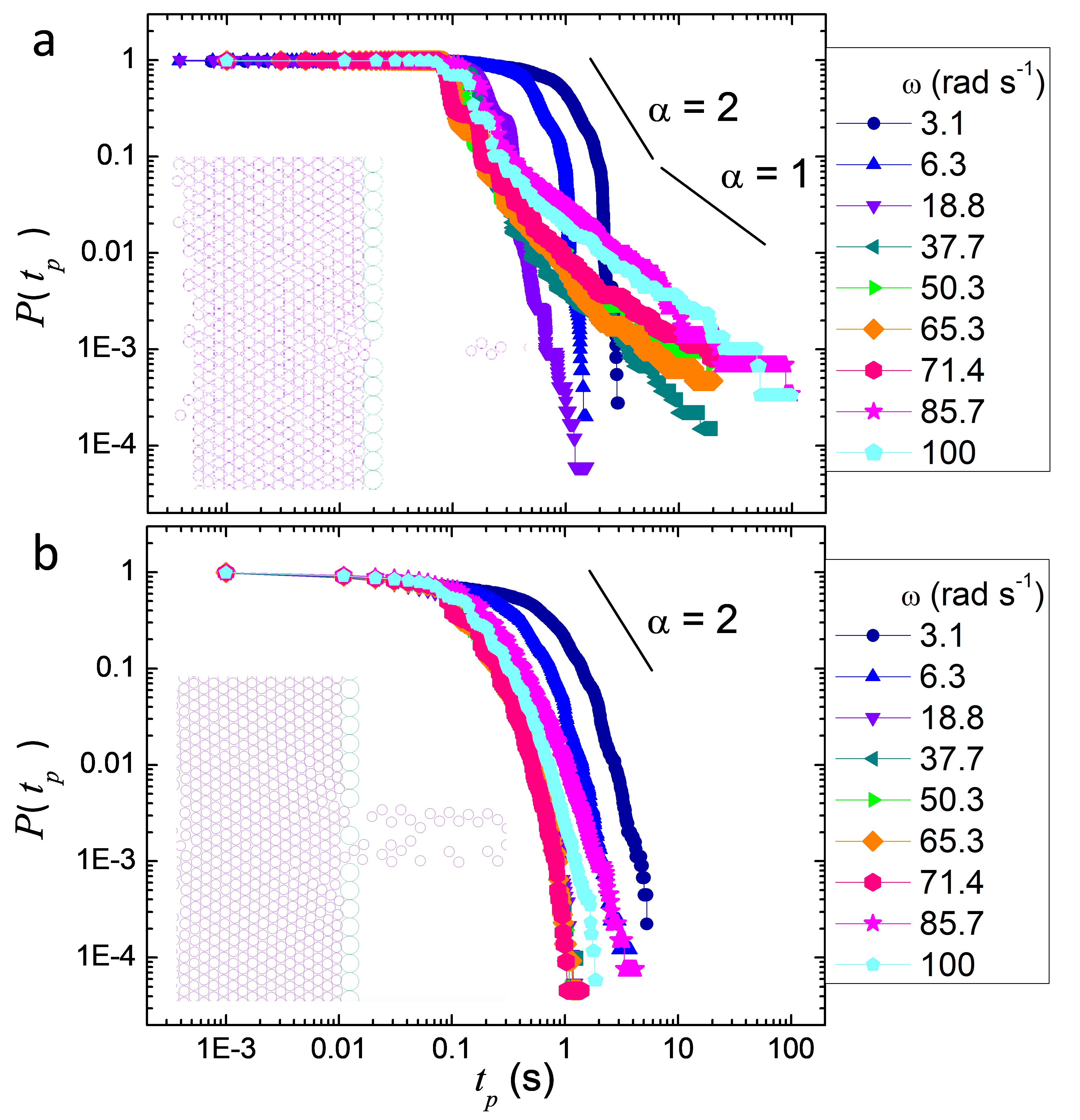}
\caption{(a,b) Distribution functions $P(t_p)$ of time lapses, $t_p$, for different frequencies of a straight wall of non magnetic colloids 
with: (a) a single constriction with width $\delta=2.982 \rm{\mu m}$ and (b), two small constrictions both with with width $\delta=2.982 \rm{\mu m}$ and separated by a single particle.  
The small insets on the left of both images show a fraction of the the simulation 
box with a close packed monolayer of particles. The inset at the top show channel clogging, while the inset at the bottom simple particle flow.}
\label{figure6}
\end{center}
\end{figure}
 
\subsection*{Simulation results: planar wall}
The geometry analyzed in the previous section with numerical simulation was designed to match the experimental system, thus featuring obstacles with the same position and degrees of disorder. Such disorder results from the procedure to prepare the obstacles, since the silica particles were left sediment above the substrate and forced to stick there by addition of salt. Thus, adapting the simulation to the experimental system 
implies a random distribution of the particles and an opening with a small inclination with respect to the driving direction, as shown in Fig.~\ref{figure1}(c).
However, it is also interesting to analyze the generic features of the probability of elapsed times on a simplified geometry, when the silica obstacles form a perfect planar wall with a small opening at the center. This geometry is depicted in the inset of  Fig.~\ref{figure6}(a), for a width $\delta = 2.982 \rm{\mu m}$, while the main figure shows the variation of  $P(t_p )$ as a function of  $\omega$.
At very low frequencies, $\omega < 18.8 \rm{rad s^{-1}}$,  
the distributions are exponential and the particles easily flow trough the aperture. Increasing $\omega$ we observe the emergence of the steps as in the previous case, however we find that  
clogging may be observed even within the synchronous regime ($\omega = 65.3 \rm{rad s^{-1}}$) where the power law distributions feature a very small exponent $\alpha \sim 1.2$. 
The reason is that, in contrast to the 
disordered geometry, the planar
may act as a template for crystallization (see inset Fig.~\ref{figure6}(a))
and create arches and bridges which easily impeded further motion, creating an intermittent flow at the opening. In contrast at low frequency the particle oscillations due to the moving landscape are able to break these bridges and to avoid the crystallization process.

Further, the numerical system allows to explore the effect that an additional opening in the straight wall will have on the global particle flow. An interesting question is whether adding a second aperture separated by only one silica obstacle, could favor the system fluidization and avoid clogging. We effectively report this effect in Fig.~\ref{figure6}(b), where two apertures of width $\delta = 2.982 \rm{\mu m}$ are able to avoid the particle crystallization which induces clogging for all driving frequencies.  The measured distribution functions $P(t_p)$ for both apertures are almost identical, and in both cases  (in Fig.~\ref{figure6}(b) we show their average) display a smooth and almost exponential decay. We note that the effect of two openings may be considered analogous to placing a single obstacle at the exit of one large aperture. This situation has been used in the past in other macroscopic systems as an efficient means to fluidize  a system prone to clogging~\cite{Zuriguel2011,Endo2017,Arean2020}. However, in most cases the obstacle was placed close to the opening, but not exactly at the exit. Another interesting avenue for future study is to investigate how the distributions function may change upon variation of different parameters, such  as the location and size of the central obstacle, or when the two opening are asymmetric, namely with different width.

\section*{Conclusions}
We have studied both experimentally and theoretically
the dynamics of paramagnetic colloidal particles driven 
through a single aperture above a periodic magnetic substrate. 
We combine experiments with numerical simulations and 
analyze the distribution of displacements of the particles. 
We find that in the experiments clogging occurs when the particles are driven 
close to the asynchronous regime, 
while in the simulations they always occur deep in this regime. Further we unveil the role played by the 
obstacle movement due to flexibility and to the hydrodynamic interactions. The comparison between experimental and simulation results on analogous geometries indicate that HI play a relevant role in the clogging dynamics of forced colloids  and that obstacle compliance  hinders the dependence of  particle  motion on the  frequency of the driving field.

While the experimental system is based on the use of a specially prepared ferrite garnet film, 
we observe a very good agreement between the experimental data and the results from numerical simulations by using a very generic model. The latter neglect dipolar interactions between the particles while considering an effective short range repulsive interaction potential. Thus our findings may be used to explore clogging in other driven many particle systems two dimensionally confined to a plate, not limited to magnetic ones.
On the other hand, the possibility of increasing 
the dipolar interactions between the particles via the ellipticity of the applied field could be further used as an effective tool to switch on attractive interactions and induce chaining. 

Moreover, our findings invite future explorations of the system, as considering two, three or several opening in different geometric arrangements.  Another interesting avenue would be to explore how the overall dynamics changes for anisotropic magnetic particles driven through the aperture.~\cite{Zerrouki2008,Lee09,Yan2012,Palacci2013,Tierno14,Martinez2018} On the application side, the possibility of transporting 
paramagnetic colloidal particles close to a surface, and localize their position by simply switching off the applied field 
may be of interest for microfluidics~\cite{Terray2002} and lab-on-a-chip~\cite{Wen2010}
systems. In particular, these particles can be used to pick up and mobilize a chemical or biological cargo via surface functionalization~\cite{Garg2015}.

\section*{Conflicts of interest}
There are no conflicts to declare.

\section*{Acknowledgements}
We thank Tom H. Johansen for providing us with the FGF film.
R. L. S. acknowledges support
from the Swiss National Science Foundation Grant 180729. R. L. S. and P. T. acknowledge support from the ERC consolidator grant  Enforce (No. 811234).   P. T. acknowledges support from the Ministerio de Ciencia, Innovaci\'on y Universidades (Grant no. ERC2018-092827) and the Generalitat de Catalunya (2017SGR1061 and program ICREA Acad\`emia).
I. P. acknowledges support from Ministerio de Ciencia, Innovaci\'on y Universidades (Grant No. PGC2018-098373-B-100), DURSI (Grant No. 2017 SGR 884), and SNF (Project No. 200021-
175719). 

\bibliography{Bibliography}

\providecommand*{\mcitethebibliography}{\thebibliography}
\csname @ifundefined\endcsname{endmcitethebibliography}
{\let\endmcitethebibliography\endthebibliography}{}
\begin{mcitethebibliography}{46}
\providecommand*{\natexlab}[1]{#1}
\providecommand*{\mciteSetBstSublistMode}[1]{}
\providecommand*{\mciteSetBstMaxWidthForm}[2]{}
\providecommand*{\mciteBstWouldAddEndPuncttrue}
  {\def\EndOfBibitem{\unskip.}}
\providecommand*{\mciteBstWouldAddEndPunctfalse}
  {\let\EndOfBibitem\relax}
\providecommand*{\mciteSetBstMidEndSepPunct}[3]{}
\providecommand*{\mciteSetBstSublistLabelBeginEnd}[3]{}
\providecommand*{\EndOfBibitem}{}
\mciteSetBstSublistMode{f}
\mciteSetBstMaxWidthForm{subitem}
{(\emph{\alph{mcitesubitemcount}})}
\mciteSetBstSublistLabelBeginEnd{\mcitemaxwidthsubitemform\space}
{\relax}{\relax}

\bibitem[Isichenko(1992)]{Isichenko1992}
M.~B. Isichenko, \emph{Rev. Mod. Phys.}, 1992, \textbf{64}, 961--1043\relax
\mciteBstWouldAddEndPuncttrue
\mciteSetBstMidEndSepPunct{\mcitedefaultmidpunct}
{\mcitedefaultendpunct}{\mcitedefaultseppunct}\relax
\EndOfBibitem
\bibitem[Snarskii \emph{et~al.}(2016)Snarskii, Bezsudnov, Sevryukov,
  Morozovskiy, and Malinsky]{Snarskii2016}
A.~A. Snarskii, I.~V. Bezsudnov, V.~A. Sevryukov, A.~Morozovskiy and
  J.~Malinsky, \emph{Transport Processes in Macroscopically Disordered Media:
  From Mean Field Theory to Percolation}, Springer, New York, 2016\relax
\mciteBstWouldAddEndPuncttrue
\mciteSetBstMidEndSepPunct{\mcitedefaultmidpunct}
{\mcitedefaultendpunct}{\mcitedefaultseppunct}\relax
\EndOfBibitem
\bibitem[Redner and Datta(2000)]{Redner2000}
S.~Redner and S.~Datta, \emph{Phys. Rev. Lett.}, 2000, \textbf{84},
  6018--6021\relax
\mciteBstWouldAddEndPuncttrue
\mciteSetBstMidEndSepPunct{\mcitedefaultmidpunct}
{\mcitedefaultendpunct}{\mcitedefaultseppunct}\relax
\EndOfBibitem
\bibitem[Roussel \emph{et~al.}(2007)Roussel, Nguyen, and Coussot]{Roussel2007}
N.~Roussel, T.~L.~H. Nguyen and P.~Coussot, \emph{Phys. Rev. Lett.}, 2007,
  \textbf{98}, 114502\relax
\mciteBstWouldAddEndPuncttrue
\mciteSetBstMidEndSepPunct{\mcitedefaultmidpunct}
{\mcitedefaultendpunct}{\mcitedefaultseppunct}\relax
\EndOfBibitem
\bibitem[MacDonald \emph{et~al.}(2003)MacDonald, Spalding, and
  Dholaki]{MacDonald2003}
M.~P. MacDonald, G.~C. Spalding and K.~Dholaki, \emph{Nature}, 2003,
  \textbf{426}, 421\relax
\mciteBstWouldAddEndPuncttrue
\mciteSetBstMidEndSepPunct{\mcitedefaultmidpunct}
{\mcitedefaultendpunct}{\mcitedefaultseppunct}\relax
\EndOfBibitem
\bibitem[Martinez-Pedrero \emph{et~al.}(2016)Martinez-Pedrero, Massana-Cid,
  Ziegler, Johansen, Straube, and Tierno]{Martinez2016}
F.~Martinez-Pedrero, H.~Massana-Cid, T.~Ziegler, T.~H. Johansen, A.~V. Straube
  and P.~Tierno, \emph{Phys. Chem. Chem. Phys.}, 2016, \textbf{18}, 26353\relax
\mciteBstWouldAddEndPuncttrue
\mciteSetBstMidEndSepPunct{\mcitedefaultmidpunct}
{\mcitedefaultendpunct}{\mcitedefaultseppunct}\relax
\EndOfBibitem
\bibitem[Squires and Quake(2005)]{Squires2005}
T.~M. Squires and S.~R. Quake, \emph{Rev. Mod. Phys.}, 2005, \textbf{77},
  977--1026\relax
\mciteBstWouldAddEndPuncttrue
\mciteSetBstMidEndSepPunct{\mcitedefaultmidpunct}
{\mcitedefaultendpunct}{\mcitedefaultseppunct}\relax
\EndOfBibitem
\bibitem[Whitesides(2006)]{Whitesides2006}
G.~Whitesides, \emph{Nature}, 2006, \textbf{442}, 368--373\relax
\mciteBstWouldAddEndPuncttrue
\mciteSetBstMidEndSepPunct{\mcitedefaultmidpunct}
{\mcitedefaultendpunct}{\mcitedefaultseppunct}\relax
\EndOfBibitem
\bibitem[Reichhardt and Reichhardt(2017)]{Reichhardt2017}
C.~Reichhardt and C.~J.~O. Reichhardt, \emph{Rep. Prog. Phys.}, 2017,
  \textbf{80}, 026501\relax
\mciteBstWouldAddEndPuncttrue
\mciteSetBstMidEndSepPunct{\mcitedefaultmidpunct}
{\mcitedefaultendpunct}{\mcitedefaultseppunct}\relax
\EndOfBibitem
\bibitem[Cohen and Jensen(1997)]{Cohen1997}
L.~F. Cohen and H.~J. Jensen, \emph{Rep. Prog. Phys.}, 1997, \textbf{60},
  1581\relax
\mciteBstWouldAddEndPuncttrue
\mciteSetBstMidEndSepPunct{\mcitedefaultmidpunct}
{\mcitedefaultendpunct}{\mcitedefaultseppunct}\relax
\EndOfBibitem
\bibitem[Rees \emph{et~al.}(2012)Rees, Totsuji, and Kono]{Rees2012}
D.~G. Rees, H.~Totsuji and K.~Kono, \emph{Phys. Rev. Lett.}, 2012,
  \textbf{108}, 176801\relax
\mciteBstWouldAddEndPuncttrue
\mciteSetBstMidEndSepPunct{\mcitedefaultmidpunct}
{\mcitedefaultendpunct}{\mcitedefaultseppunct}\relax
\EndOfBibitem
\bibitem[Bechinger \emph{et~al.}(2016)Bechinger, Di~Leonardo, L\"owen,
  Reichhardt, Volpe, and Volpe]{Bechinger2016}
C.~Bechinger, R.~Di~Leonardo, H.~L\"owen, C.~Reichhardt, G.~Volpe and G.~Volpe,
  \emph{Rev. Mod. Phys.}, 2016, \textbf{88}, 045006\relax
\mciteBstWouldAddEndPuncttrue
\mciteSetBstMidEndSepPunct{\mcitedefaultmidpunct}
{\mcitedefaultendpunct}{\mcitedefaultseppunct}\relax
\EndOfBibitem
\bibitem[Parisi \emph{et~al.}(2018)Parisi, Hidalgo, and Zuriguel]{Parisi2018}
D.~R. Parisi, R.~C. Hidalgo and I.~Zuriguel, \emph{Scientific Reports}, 2018,
  \textbf{8}, 9133\relax
\mciteBstWouldAddEndPuncttrue
\mciteSetBstMidEndSepPunct{\mcitedefaultmidpunct}
{\mcitedefaultendpunct}{\mcitedefaultseppunct}\relax
\EndOfBibitem
\bibitem[Reichhardt \emph{et~al.}(2015)Reichhardt, Ray, and
  Reichhardt]{Reichhardt2015}
C.~Reichhardt, D.~Ray and C.~J.~O. Reichhardt, \emph{Phys. Rev. Lett.}, 2015,
  \textbf{114}, 217202\relax
\mciteBstWouldAddEndPuncttrue
\mciteSetBstMidEndSepPunct{\mcitedefaultmidpunct}
{\mcitedefaultendpunct}{\mcitedefaultseppunct}\relax
\EndOfBibitem
\bibitem[Lozano \emph{et~al.}(2012)Lozano, Lumay, Zuriguel, Hidalgo, and
  Garcimart\'{\i}n]{Loz12}
C.~Lozano, G.~Lumay, I.~Zuriguel, R.~C. Hidalgo and A.~Garcimart\'{\i}n,
  \emph{Phys. Rev. Lett.}, 2012, \textbf{109}, 068001\relax
\mciteBstWouldAddEndPuncttrue
\mciteSetBstMidEndSepPunct{\mcitedefaultmidpunct}
{\mcitedefaultendpunct}{\mcitedefaultseppunct}\relax
\EndOfBibitem
\bibitem[Zuriguel \emph{et~al.}(2015)Zuriguel, Parisi, Hidalgo, Lozano, Janda,
  Gago, Peralta, Ferrer, Pugnaloni, Cl\'ement, Maza, Pagonabarraga, and
  Garcimart\'in]{Zuriguel2015}
I.~Zuriguel, D.~R. Parisi, R.~C. Hidalgo, C.~Lozano, A.~Janda, P.~A. Gago,
  J.~P. Peralta, L.~M. Ferrer, L.~A. Pugnaloni, E.~Cl\'ement, D.~Maza,
  I.~Pagonabarraga and A.~Garcimart\'in, \emph{Scientific Reports}, 2015,
  \textbf{4}, 7324\relax
\mciteBstWouldAddEndPuncttrue
\mciteSetBstMidEndSepPunct{\mcitedefaultmidpunct}
{\mcitedefaultendpunct}{\mcitedefaultseppunct}\relax
\EndOfBibitem
\bibitem[Helbing and Moln\'ar(1995)]{Helbing1995}
D.~Helbing and P.~Moln\'ar, \emph{Phys. Rev. E}, 1995, \textbf{51},
  4282--4286\relax
\mciteBstWouldAddEndPuncttrue
\mciteSetBstMidEndSepPunct{\mcitedefaultmidpunct}
{\mcitedefaultendpunct}{\mcitedefaultseppunct}\relax
\EndOfBibitem
\bibitem[Garcimart\'{\i}n \emph{et~al.}(2015)Garcimart\'{\i}n, Pastor, Ferrer,
  Ramos, Mart\'{\i}n-G\'omez, and Zuriguel]{Garcimartin2015}
A.~Garcimart\'{\i}n, J.~M. Pastor, L.~M. Ferrer, J.~J. Ramos,
  C.~Mart\'{\i}n-G\'omez and I.~Zuriguel, \emph{Phys. Rev. E}, 2015,
  \textbf{91}, 022808\relax
\mciteBstWouldAddEndPuncttrue
\mciteSetBstMidEndSepPunct{\mcitedefaultmidpunct}
{\mcitedefaultendpunct}{\mcitedefaultseppunct}\relax
\EndOfBibitem
\bibitem[Janda \emph{et~al.}(2009)Janda, D.Maza, Garcimart, Kolb, Lanuza, and
  Cl\'ement]{Jan09}
A.~Janda, D.Maza, A.~Garcimart, E.~Kolb, J.~Lanuza and E.~Cl\'ement,
  \emph{Europhys. Lett.}, 2009, \textbf{87}, 24002\relax
\mciteBstWouldAddEndPuncttrue
\mciteSetBstMidEndSepPunct{\mcitedefaultmidpunct}
{\mcitedefaultendpunct}{\mcitedefaultseppunct}\relax
\EndOfBibitem
\bibitem[To \emph{et~al.}(2001)To, Lai, and Pak]{To01}
K.~To, P.-Y. Lai and H.~K. Pak, \emph{Phys. Rev. Lett.}, 2001, \textbf{86},
  71--74\relax
\mciteBstWouldAddEndPuncttrue
\mciteSetBstMidEndSepPunct{\mcitedefaultmidpunct}
{\mcitedefaultendpunct}{\mcitedefaultseppunct}\relax
\EndOfBibitem
\bibitem[Galajda \emph{et~al.}(2007)Galajda, Keymer, Chaikin, and
  Austin]{Galajda8704}
P.~Galajda, J.~Keymer, P.~Chaikin and R.~Austin, \emph{Journal of
  Bacteriology}, 2007, \textbf{189}, 8704--8707\relax
\mciteBstWouldAddEndPuncttrue
\mciteSetBstMidEndSepPunct{\mcitedefaultmidpunct}
{\mcitedefaultendpunct}{\mcitedefaultseppunct}\relax
\EndOfBibitem
\bibitem[Koumakis \emph{et~al.}(2013)Koumakis, Lepore, Maggi, and
  Di~Leonardo]{Koumakis2013}
N.~Koumakis, A.~Lepore, C.~Maggi and R.~Di~Leonardo, \emph{Nat. Comm.}, 2013,
  \textbf{4}, 2588\relax
\mciteBstWouldAddEndPuncttrue
\mciteSetBstMidEndSepPunct{\mcitedefaultmidpunct}
{\mcitedefaultendpunct}{\mcitedefaultseppunct}\relax
\EndOfBibitem
\bibitem[Reichhardt and Reichhardt(2018)]{Reichhardt2018}
C.~Reichhardt and C.~J.~O. Reichhardt, \emph{Phys. Rev. Lett.}, 2018,
  \textbf{121}, 068001\relax
\mciteBstWouldAddEndPuncttrue
\mciteSetBstMidEndSepPunct{\mcitedefaultmidpunct}
{\mcitedefaultendpunct}{\mcitedefaultseppunct}\relax
\EndOfBibitem
\bibitem[Haw(2004)]{Haw2004}
M.~D. Haw, \emph{Phys. Rev. Lett.}, 2004, \textbf{92}, 185506\relax
\mciteBstWouldAddEndPuncttrue
\mciteSetBstMidEndSepPunct{\mcitedefaultmidpunct}
{\mcitedefaultendpunct}{\mcitedefaultseppunct}\relax
\EndOfBibitem
\bibitem[Marin \emph{et~al.}(2018)Marin, Lhuissier, Rossi, and
  K\"ahler]{Marin2018}
A.~Marin, H.~Lhuissier, M.~Rossi and C.~J. K\"ahler, \emph{Phys. Rev. E}, 2018,
  \textbf{97}, 021102\relax
\mciteBstWouldAddEndPuncttrue
\mciteSetBstMidEndSepPunct{\mcitedefaultmidpunct}
{\mcitedefaultendpunct}{\mcitedefaultseppunct}\relax
\EndOfBibitem
\bibitem[Stoop and Tierno(2018)]{Stoop18}
R.~L. Stoop and P.~Tierno, \emph{Communications Phys.}, 2018, \textbf{1},
  68\relax
\mciteBstWouldAddEndPuncttrue
\mciteSetBstMidEndSepPunct{\mcitedefaultmidpunct}
{\mcitedefaultendpunct}{\mcitedefaultseppunct}\relax
\EndOfBibitem
\bibitem[Tierno \emph{et~al.}(2009)Tierno, Sagués, Johansen, and
  Fischer]{Tierno2009}
P.~Tierno, F.~Sagués, T.~H. Johansen and T.~M. Fischer, \emph{Phys. Chem.
  Chem. Phys.}, 2009, \textbf{11}, 9615--9625\relax
\mciteBstWouldAddEndPuncttrue
\mciteSetBstMidEndSepPunct{\mcitedefaultmidpunct}
{\mcitedefaultendpunct}{\mcitedefaultseppunct}\relax
\EndOfBibitem
\bibitem[Tierno(2012)]{Tierno2012}
P.~Tierno, \emph{Phys. Rev. Lett.}, 2012, \textbf{109}, 198304\relax
\mciteBstWouldAddEndPuncttrue
\mciteSetBstMidEndSepPunct{\mcitedefaultmidpunct}
{\mcitedefaultendpunct}{\mcitedefaultseppunct}\relax
\EndOfBibitem
\bibitem[Straube and Tierno(2014)]{Str14}
A.~V. Straube and P.~Tierno, \emph{Soft Matter}, 2014, \textbf{10}, 3915\relax
\mciteBstWouldAddEndPuncttrue
\mciteSetBstMidEndSepPunct{\mcitedefaultmidpunct}
{\mcitedefaultendpunct}{\mcitedefaultseppunct}\relax
\EndOfBibitem
\bibitem[Straube and Tierno(2013)]{Str13}
A.~V. Straube and P.~Tierno, \emph{Europhys. Lett.}, 2013, \textbf{103},
  28001\relax
\mciteBstWouldAddEndPuncttrue
\mciteSetBstMidEndSepPunct{\mcitedefaultmidpunct}
{\mcitedefaultendpunct}{\mcitedefaultseppunct}\relax
\EndOfBibitem
\bibitem[Tierno \emph{et~al.}(2007)Tierno, Muruganathan, and
  Fischer]{Tierno2007}
P.~Tierno, R.~Muruganathan and T.~M. Fischer, \emph{Phys. Rev. Lett.}, 2007,
  \textbf{98}, 028301\relax
\mciteBstWouldAddEndPuncttrue
\mciteSetBstMidEndSepPunct{\mcitedefaultmidpunct}
{\mcitedefaultendpunct}{\mcitedefaultseppunct}\relax
\EndOfBibitem
\bibitem[Blake and Chwang(1974)]{Blake}
J.~R. Blake and A.~T.~J. Chwang, \emph{Eng Math}, 1974, \textbf{8}, 23\relax
\mciteBstWouldAddEndPuncttrue
\mciteSetBstMidEndSepPunct{\mcitedefaultmidpunct}
{\mcitedefaultendpunct}{\mcitedefaultseppunct}\relax
\EndOfBibitem
\bibitem[Saloma \emph{et~al.}(2003)Saloma, Perez, Tapang, Lim, and
  Palmes-Saloma]{Saloma2003}
C.~Saloma, G.~J. Perez, G.~Tapang, M.~Lim and C.~Palmes-Saloma, \emph{Proc.
  Nat. Acad. Sci. USA}, 2003, \textbf{100}, 11947\relax
\mciteBstWouldAddEndPuncttrue
\mciteSetBstMidEndSepPunct{\mcitedefaultmidpunct}
{\mcitedefaultendpunct}{\mcitedefaultseppunct}\relax
\EndOfBibitem
\bibitem[Clauset \emph{et~al.}(2009)Clauset, Shalizi, and Newman]{Cla09}
A.~Clauset, C.~R. Shalizi and M.~E.~J. Newman, \emph{SIAM review}, 2009,
  \textbf{51}, 661\relax
\mciteBstWouldAddEndPuncttrue
\mciteSetBstMidEndSepPunct{\mcitedefaultmidpunct}
{\mcitedefaultendpunct}{\mcitedefaultseppunct}\relax
\EndOfBibitem
\bibitem[Zuriguel \emph{et~al.}(2011)Zuriguel, Janda, Garcimart\'{\i}n, Lozano,
  Ar\'evalo, and Maza]{Zuriguel2011}
I.~Zuriguel, A.~Janda, A.~Garcimart\'{\i}n, C.~Lozano, R.~Ar\'evalo and
  D.~Maza, \emph{Phys. Rev. Lett.}, 2011, \textbf{107}, 278001\relax
\mciteBstWouldAddEndPuncttrue
\mciteSetBstMidEndSepPunct{\mcitedefaultmidpunct}
{\mcitedefaultendpunct}{\mcitedefaultseppunct}\relax
\EndOfBibitem
\bibitem[Endo \emph{et~al.}(2017)Endo, Reddy, and Katsuragi]{Endo2017}
K.~Endo, K.~A. Reddy and H.~Katsuragi, \emph{Phys. Rev. Fluids}, 2017,
  \textbf{2}, 094302\relax
\mciteBstWouldAddEndPuncttrue
\mciteSetBstMidEndSepPunct{\mcitedefaultmidpunct}
{\mcitedefaultendpunct}{\mcitedefaultseppunct}\relax
\EndOfBibitem
\bibitem[Are\'an \emph{et~al.}(2020)Are\'an, Boschan, Cachile, and
  Aguirre]{Arean2020}
M.~G. Are\'an, A.~Boschan, M.~A. Cachile and M.~A. Aguirre, \emph{Phys. Rev.
  E}, 2020, \textbf{101}, 022901\relax
\mciteBstWouldAddEndPuncttrue
\mciteSetBstMidEndSepPunct{\mcitedefaultmidpunct}
{\mcitedefaultendpunct}{\mcitedefaultseppunct}\relax
\EndOfBibitem
\bibitem[Zerrouki \emph{et~al.}(2008)Zerrouki, Baudry, Pine, Chaikin, and
  Bibette]{Zerrouki2008}
D.~Zerrouki, J.~Baudry, D.~Pine, P.~Chaikin and J.~Bibette, \emph{Nature},
  2008, \textbf{455}, 380\relax
\mciteBstWouldAddEndPuncttrue
\mciteSetBstMidEndSepPunct{\mcitedefaultmidpunct}
{\mcitedefaultendpunct}{\mcitedefaultseppunct}\relax
\EndOfBibitem
\bibitem[H. and M.(2009)]{Lee09}
L.~S. H. and L.~C. M., \emph{Small}, 2009, \textbf{5}, 1957\relax
\mciteBstWouldAddEndPuncttrue
\mciteSetBstMidEndSepPunct{\mcitedefaultmidpunct}
{\mcitedefaultendpunct}{\mcitedefaultseppunct}\relax
\EndOfBibitem
\bibitem[Yan \emph{et~al.}(2012)Yan, Bloom, Bae, E., and Granick]{Yan2012}
J.~Yan, M.~Bloom, S.~C. Bae, L.~E. and S.~Granick, \emph{Nature}, 2012,
  \textbf{491}, 578\relax
\mciteBstWouldAddEndPuncttrue
\mciteSetBstMidEndSepPunct{\mcitedefaultmidpunct}
{\mcitedefaultendpunct}{\mcitedefaultseppunct}\relax
\EndOfBibitem
\bibitem[Palacci \emph{et~al.}(2013)Palacci, Sacanna, Vatchinsky, Chaikin, and
  Pine]{Palacci2013}
J.~Palacci, S.~Sacanna, A.~Vatchinsky, P.~M. Chaikin and D.~J. Pine, \emph{J.
  Am. Chem. Soc.}, 2013, \textbf{135}, 15978\relax
\mciteBstWouldAddEndPuncttrue
\mciteSetBstMidEndSepPunct{\mcitedefaultmidpunct}
{\mcitedefaultendpunct}{\mcitedefaultseppunct}\relax
\EndOfBibitem
\bibitem[Tierno(2014)]{Tierno14}
P.~Tierno, \emph{Phys. Chem. Chem. Phys.}, 2014, \textbf{16}, 23515\relax
\mciteBstWouldAddEndPuncttrue
\mciteSetBstMidEndSepPunct{\mcitedefaultmidpunct}
{\mcitedefaultendpunct}{\mcitedefaultseppunct}\relax
\EndOfBibitem
\bibitem[Martinez-Pedrero \emph{et~al.}(2018)Martinez-Pedrero,
  Navarro-Argem\'i, Ortiz-Ambriz, Pagonabarraga, and Tierno]{Martinez2018}
F.~Martinez-Pedrero, E.~Navarro-Argem\'i, A.~Ortiz-Ambriz, I.~Pagonabarraga and
  P.~Tierno, \emph{Science Adv.}, 2018, \textbf{4}, 9379\relax
\mciteBstWouldAddEndPuncttrue
\mciteSetBstMidEndSepPunct{\mcitedefaultmidpunct}
{\mcitedefaultendpunct}{\mcitedefaultseppunct}\relax
\EndOfBibitem
\bibitem[Terray \emph{et~al.}(2002)Terray, Oakey, and Marr]{Terray2002}
A.~Terray, J.~Oakey and D.~W. Marr, \emph{Science}, 2002, \textbf{296},
  1841\relax
\mciteBstWouldAddEndPuncttrue
\mciteSetBstMidEndSepPunct{\mcitedefaultmidpunct}
{\mcitedefaultendpunct}{\mcitedefaultseppunct}\relax
\EndOfBibitem
\bibitem[Wang \emph{et~al.}(2010)Wang, Zhang, Li, Gong, and Wen]{Wen2010}
L.~Wang, M.~Zhang, J.~Li, X.~Gong and W.~Wen, \emph{Lab chip}, 2010,
  \textbf{10}, 2869\relax
\mciteBstWouldAddEndPuncttrue
\mciteSetBstMidEndSepPunct{\mcitedefaultmidpunct}
{\mcitedefaultendpunct}{\mcitedefaultseppunct}\relax
\EndOfBibitem
\bibitem[Garg \emph{et~al.}(2015)Garg, Rath, and Goyal]{Garg2015}
T.~Garg, G.~Rath and A.~K. Goyal, \emph{Crit. Rev. Ther. Drug Carrier Syst.},
  2015, \textbf{32}, 89\relax
\mciteBstWouldAddEndPuncttrue
\mciteSetBstMidEndSepPunct{\mcitedefaultmidpunct}
{\mcitedefaultendpunct}{\mcitedefaultseppunct}\relax
\EndOfBibitem
\end{mcitethebibliography}
\bibliographystyle{rsc} 
\end{document}